\documentclass[doublecol]{epl2}
\usepackage{tabularx}% Table width

\title{Superconductivity at 2.5 K in new transition-metal chalcogenide Ta$_2$PdSe$_5$}
\shorttitle{} %Insert here a short version of the title if it exceeds 70 characters

\author{J. Zhang$^1$, J. K. Dong$^{1,2}$\footnote{E-mail: jkdong@fudan.edu.cn}, Y. Xu$^1$, J. Pan$^1$, L. P. He$^1$, L. J. Zhang$^2$, and S. Y. Li$^{1,2,3}$\footnote{E-mail: shiyan$\_$li@fudan.edu.cn}}
\shortauthor{J. Zhang\etal}

\institute{$^1$State Key Laboratory of Surface Physics and Department of Physics, Fudan University, Shanghai 200433, P. R. China\\
$^2$Laboratory of Advanced Materials, Fudan University, Shanghai 200438, P. R. China\\
$^3$Collaborative Innovation Center of Advanced Microstructures, Fudan University, Shanghai 200433, P. R. China}

	{Transport properties}

\abstract{We report the synthesis and superconducting properties of a new transition-metal chalcogenide Ta$_2$PdSe$_5$. The measurements of resistivity, magnetization, and specific heat reveal that Ta$_2$PdSe$_5$ is a bulk superconductor with $T_c$ $\simeq$ 2.5 K. The zero-field electronic specific heat in the superconducting state can be fitted with a two-gap BCS model. The upper critical field $H_{c2}$ shows a linear temperature dependence, and the value of $H_{c2}$(0) is much higher than the estimated Pauli limiting field $H_{c2}^{P}$ and orbital limiting field $H_{c2}^{orb}$. All these results of specific heat and upper critical field suggest that Ta$_2$PdSe$_5$ is a multi-band superconductor.}

\begin{document}

\maketitle
\section{Introduction}
For BCS superconductors with spin-singlet pairing, magnetic field destroys the superconductivity by either orbital effect or Pauli paramagnetic effect.
The upper critical field $H_{c2}$ which reflects the pair-breaking mechanism is usually limited by the orbital effect, with the limiting field $H_{c2}^{orb}$ = $\Phi_0$/2$\pi\xi^2$.
Pauli paramagnetic effect becomes important when the orbital contribution is somehow suppressed or when the normal-state spin susceptibility $\chi_N$ is enhanced due to spin-orbit coupling, resulting in the Pauli limiting field $H_{c2}^{P}$ smaller than $H_{c2}^{orb}$ \cite{Matsuda}. Then the spin polarization determines the upper critical field $H_{c2}^{P} =$ 1.84 $T_c$, within the weak-coupling BCS theory \cite{Clogston}.
Experimentally, superconductivity with $H_{c2}$ beyond $H_{c2}^{P}$ has been observed in some low-dimensional materials, such as the organic Bechgaard salt (TMTSF)$_2$PF$_6$ \cite{Lee}, purple bronze Li$_{0.9}$Mo$_6$O$_{17}$ \cite{Mercure}, and iron chalcogenide $\beta$-FeSe \cite{Lei}.
Different theories including spin-triplet pairing, strong spin-orbit coupling, and multi-band effect are proposed \cite{Lei,Mercure,Lee}.
For practical applications, the superconductors with high $H_{c2}$ are more promising in high-field application.

Recently, a new quasi-one-dimentional (Q1D) transition-metal chalcogenide Nb$_2$Pd$_{0.81}$S$_5$ was synthesized, in the monoclinic space group $C2/m$ \cite{Zhang}. It becomes a superconductor below the transition temperature $T_c \simeq$ 6.6 K \cite{Zhang}. Later, the two existing compounds Ta$_2$PdS$_5$ and Nb$_2$PdSe$_5$ with the same crystal structure \cite{Squattrito}, were also found to be superconducting below 6 K and 5.5 K, respectively \cite{Khim,Lu}. All these three compounds display extremely high and anisotropic $H_{c2}$ \cite{Zhang,Khim,Lu,Yu,Jha}, suggesting a new family of exotic superconductors $T_2$Pd$Ch_5$ ($T$ = Nb or Ta, $Ch$ = S or Se).

Unconventional spin-triplet pairing state has been discussed based on the quasi-one dimensionality of $T_2$Pd$Ch_5$, which may explain the unusual $H_{c2}$ \cite{Lu,Khim,Zhang}. However, this proposal is challenged by the robustness of the $H_{c2}$/$T_c$ ratio in the Se substitution experiment in Nb$_2$PdS$_5$ \cite{Niu}. The spin-orbit coupling was also considered as one possible origin of the high $H_{c2}$ in this family \cite{Lu,Khim}. This proposal is supported by the opposite substituting effect on the $H_{c2}$/$T_c$ ratio in Nb$_2$PdS$_5$ \cite{Zhou}. The substitution of Pd by heavier Pt enhances the $H_{c2}$/$T_c$ ratio, while lighter Ni substitution decreases it \cite{Zhou}. Alternatively, multi-band effect could also give rise to their high $H_{c2}$ values \cite{Zhang,Khim,Niu,Zhou}. Electronic structure calculations have shown that all these three compounds are multi-band superconductors with several sheets of Fermi surface \cite{Zhang,Singh,Khim}. Since the origin of the enhanced $H_{c2}$ is still not clear, finding more compounds in this family may provide us better understanding of their unusual $H_{c2}$/$T_c$.

In this letter, we report the synthesis and superconducting properties of a new transition-metal chalcogenide Ta$_2$PdSe$_5$, the fourth member of $T_2$Pd$Ch_5$ family. Measurements of electrical resistivity, magnetization and specific heat confirm bulk superconductivity in this material, with $T_c$ $\simeq$ 2.5 K. Both the fit of electronic specific heat and the linear temperature dependence of $H_{c2}(T)$ support multi-band superconductivity in Ta$_2$PdSe$_5$. The ratio of $H_{c2}$/$T_c$ is as high as 6 T/K, which may also be ascribed to the multi-band effect.

\section{Experimental details}

The polycrystalline samples of Ta$_2$PdSe$_5$ were synthesized by a conventional solid state reaction method. The starting materials of Ta(99.99$\%$), Pd(99.99$\%$), and Se(99.99$\%$) powders were mixed thoroughly in the ratio of 2:1:5.5 and pressed into pellets in a argon-filled glove box. Excess amount of Se is necessary to compensate the loss of Se during the reaction due to its high vapor pressure. The pellets were placed in a quartz tube, slowly heated to 750$^{\circ}$C at a rate of 10$^{\circ}$C/h and then kept at this temperature for 48 hours before shutting down the furnace. This process was repeated four times, and the finally obtained samples are stable in air. X-ray diffraction (XRD) measurement was performed by using an X-ray diffractometer (D8 Advance, Bruker) with Cu $K\alpha$ radiation. The dc magnetization was measured in a Superconducting Quantum Interference Device (SQUID, Quantum Design). Electrical resistivity measurements were performed in $^4$He and $^3$He cryostats, by a standard four-probe technique. The low-temperature specific heat was measured from 0.3 to 10 K in a Physical Property Measurement System (PPMS, Quantum Design) equipped with a small dilution refrigerator.

\section{Results and Discussion}
\begin{figure}
\includegraphics[clip,width=8cm]{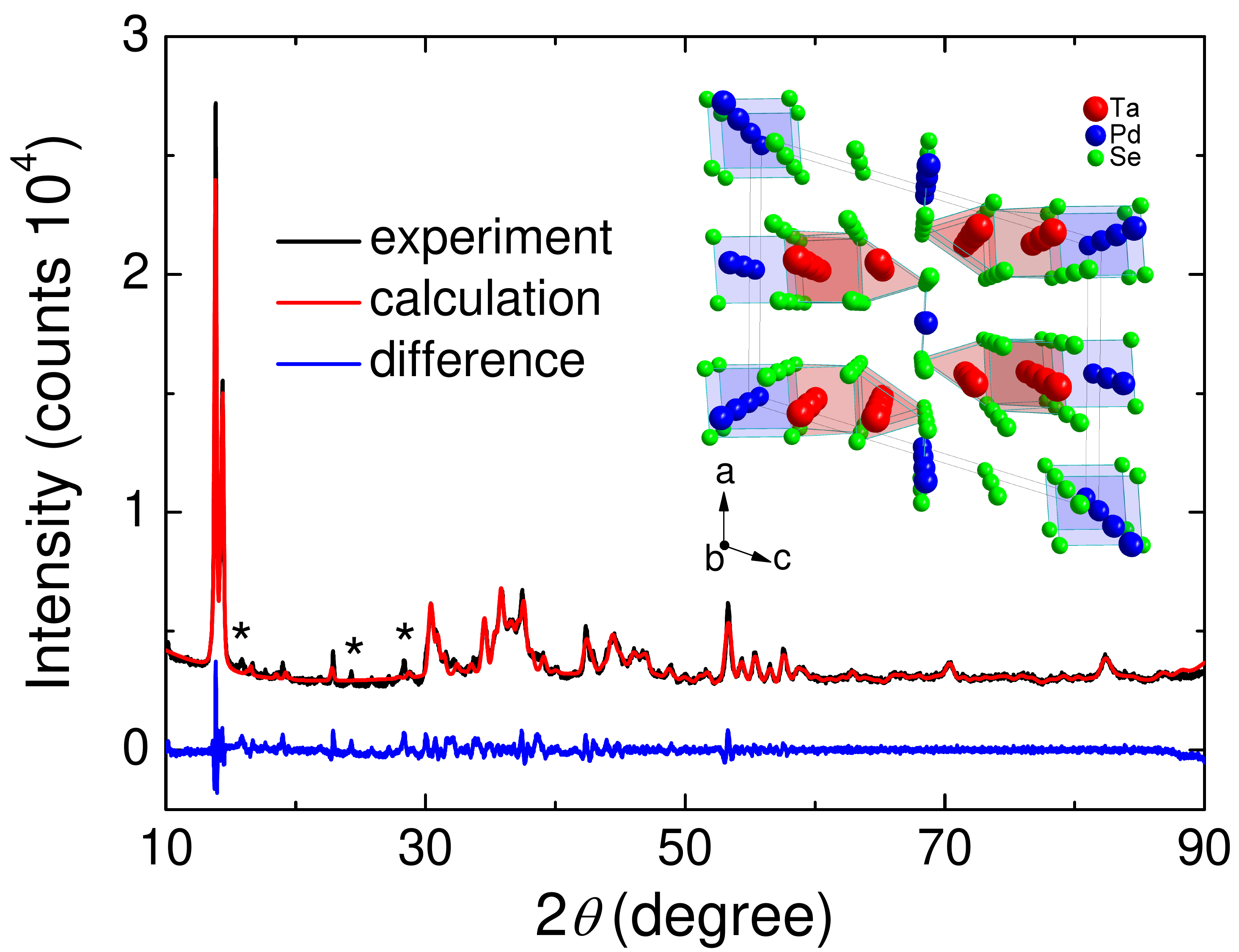}
\caption{(Color online) X-ray diffraction pattern of a Ta$_2$PdSe$_5$ polycrystalline sample measured at room temperature. The black line corresponds to the experimental XRD data, and the red line is a Rietveld refinement fit with $C2/m$ space group. The blue line is their difference. The refined
lattice parameters are $a$ = 12.801(5) {\AA}, $b$ = 3.117(1) {\AA}, $c$ = 17.742(3) {\AA},  $\alpha$ = $\gamma$ = 90$^\circ$, and $\beta$ = 106.22(2)$^\circ$. The asterisks mark the peaks of a small amount of Pd$_7$Se$_4$ impurity. The inset shows the crystal structure of Ta$_2$PdSe$_5$, an perspective view along the $b$-axis direction.}
\end{figure}

Figure 1 shows the powder X-ray diffraction (XRD) pattern of Ta$_2$PdSe$_5$ taken at room temperature. Most of the peaks can be well indexed to a monoclinic structure with space group $C2/m$. The observed XRD pattern is refined by a Rietveld method using TOPAS Academic package, from which Ta$_2$PdSe$_5$ is recognized as the main phase, with a small amount of Pd$_7$Se$_4$ impurity. The refined lattice parameters of Ta$_2$PdSe$_5$ are $a$ = 12.801(5) {\AA}, $b$ = 3.117(1) {\AA}, $c$ = 17.742(3) {\AA}, $\alpha$ = $\gamma$ = 90$^\circ$, and $\beta$ = 106.22(2)$^\circ$. A perspective view of the Ta$_2$PdSe$_5$ crystal structure along $b$-axis direction is shown in the inset of Fig. 1. The one-dimensional Ta-Se chains along the $b$ axis are bridged by PdSe$_4$ squares. This compound is the fourth member of $T_2$Pd$Ch_5$ ($T$ = Nb or Ta, $Ch$ = S or Se) family.

Figure 2(a) plots the low-temperature dc magnetization of Ta$_2$PdSe$_5$ in $H$ = 10 Oe, with both zero-field-cooled (ZFC) and field-cooled (FC) processes. The diamagnetic signal reveals a superconducting transition with the onset $T_c$ at about 2.6 K. Figure 2(b) shows the temperature dependence of resistivity for Ta$_2$PdSe$_5$ measured between 1.5 and 300 K. $\rho(T)$ displays metallic behavior with a residual resistivity ratio RRR = $\rho$(294 K)/$\rho$(3 K) $\simeq$ 3.3. The inset shows the resistivity at low temperature ranging from 1 to 5 K. A clear drop of resistivity is observed, corresponding to the superconducting transition. The onset, mid-point and zero-point temperatures of the resistive transition ($T_c^{onset}$, $T_c^{mid}$ and $T_c^{zero}$) are 2.5, 2.2, and 2.0 K, respectively. The $T_c^{onset}$ is determined from the intersection of the two extrapolated lines near the transition, as seen in the inset. The $T_c^{mid}$ and $T_c^{zero}$ are defined at the temperatures where the normal-state resistivity drops by half and to zero, respectively. The width of the resistive superconducting transition $\Delta$$T_c$ is about 0.5 K.

\begin{figure}
\includegraphics[clip,width=8cm]{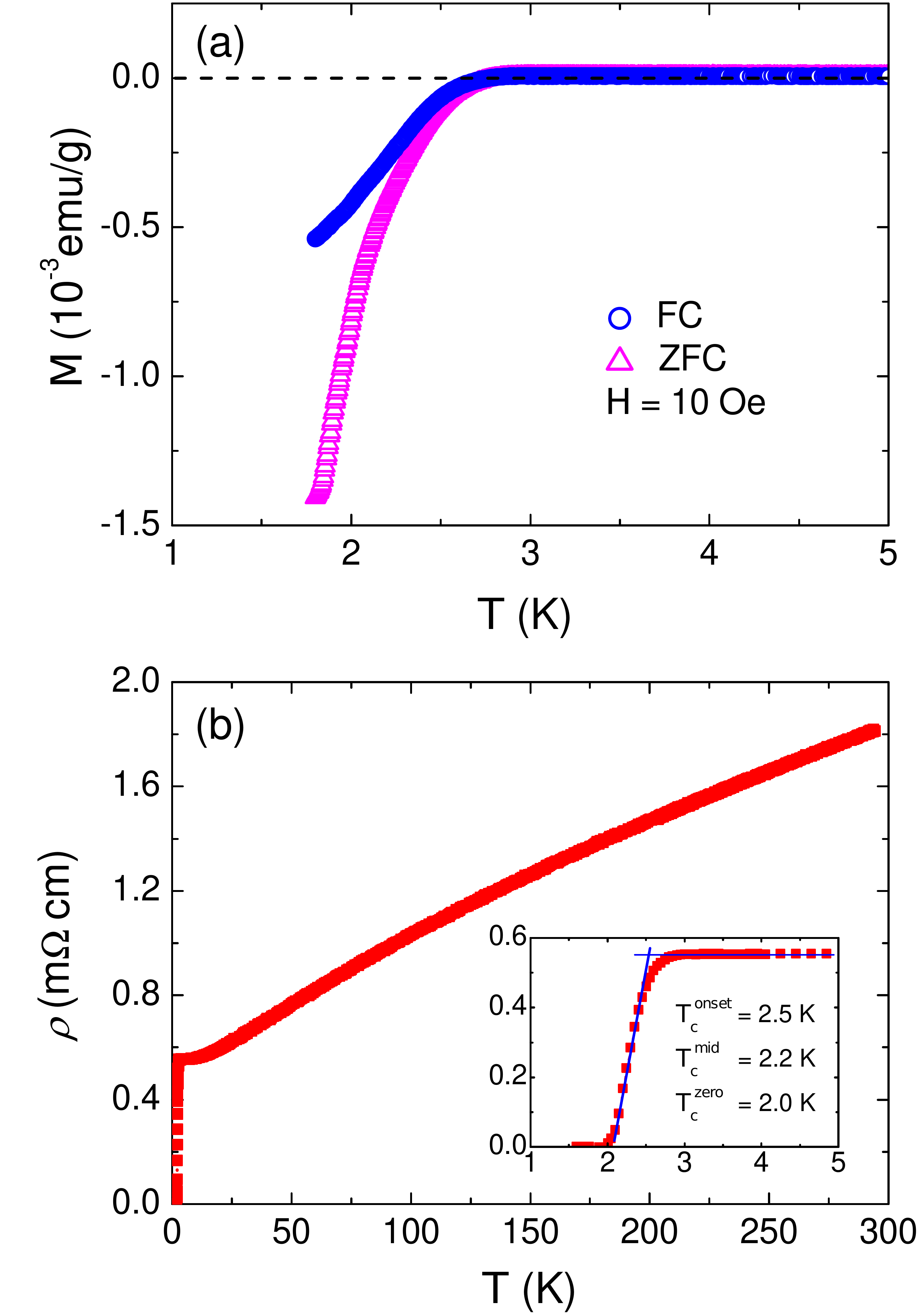}
\caption{(Color online) (a) Low-temperature dc magnetization of Ta$_2$PdSe$_5$ measured with both field-cooled (FC) and zero-field-cooled (ZFC) processes. The superconducting transition is observed at about 2.6 K. (b) Temperature dependence of the resistivity $\rho(T)$. The inset shows the superconducting transition at low temperature, with different definitions of the transition temperature $T_c$.}
\end{figure}

Figure 3(a) plots the specific heat divided by the temperature, $C_p/T$, as a function of temperature in various magnetic fields. $C_p/T$ shows an anomaly around 2.5 K in zero field, corresponding to the superconducting transition. This anomaly shifts to lower temperature and becomes less pronounced with increasing magnetic field, as shown in the inset. Above $T_c$, the zero-field data can be well fitted by $C_p/T$ = $\gamma$ + $\beta$$T^2$ + $\delta$$T^4$ from 3.5 to 10 K. We determine the electronic specific heat coefficient $\gamma$, the phononic coefficients $\beta$ to be 10.3 mJ mol$^{-1}$ K$^{-2}$ and 3.2 mJ mol$^{-1}$ K$^{-4}$, respectively. The Debye temperature $\Theta_D$ $\simeq$ 169 K is estimated from the equation $\beta$ = (12$\pi^4$$nk_B$)/(5$\Theta_D^3$), where $n$ is the number of atoms per formula unit.

\begin{figure}
\includegraphics[clip,width=8.4cm]{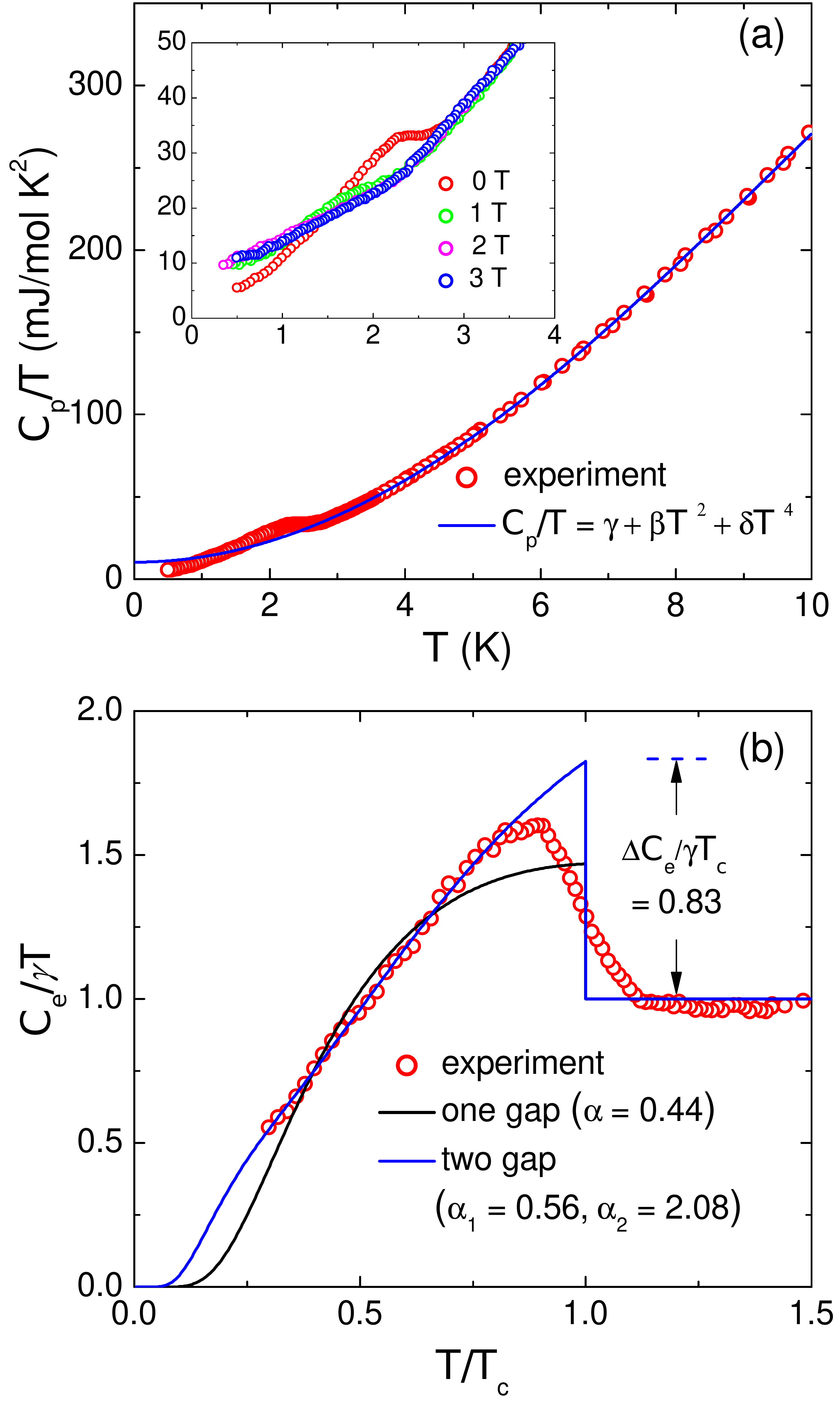}
\caption{(Color online) (a) Temperature dependence of specific heat divided by temperature $C_p/T$ in zero field. The blue line is a fit to equation $C_p/T$ = $\gamma$ + $\beta$$T^2$ + $\delta$$T^4$. Inset: temperature dependence of $C_p/T$ near the superconducting transition, measured at $H$ = 0, 1, 2, and 3 T.
 (b) Reduced temperature $T/T_c$ dependence of electronic specific heat divided by temperature $C_{e}/T$.}
\end{figure}

The zero-field electronic specific heat $C_e/T$ obtained by subtracting the lattice terms from $C_p$ is depicted in Fig. 3(b).
The bulk nature of the superconductivity in Ta$_2$PdSe$_5$ is confirmed by the significant jump of $C_e/T$.
In order to get more information about the superconducting gap, we fit $C_e/T$ in the superconducting state with the BCS $\alpha$-model $C_e$ = $C_0$exp(-$\Delta/k_BT$), where $\Delta$ is the size of the superconducting gap. It is found that the one-gap model with $\alpha$ $\equiv$ $\Delta/k_BT_c$ = 0.44 can not describe the experimental data well, while the two-gap model with $\alpha_1$ = 0.56 and $\alpha_2$ = 2.08 gives the best fit \cite{Padamsee}. The superconducting gap ratio $\Delta_l$/$\Delta_s$ $\simeq$ 3.7 is obtained. This result is consistent with the band structure calculation of $T_2$Pd$Ch_5$ \cite{Zhang,Khim,Singh}. Taking spin-orbit coupling into account, the calculated Fermi surface of Ta$_2$PdS$_5$ consists of a large hole sheet and two electron sheets (a large two dimensional cylinder around the zone center and small closed sections on the zone faces), and the superconductivity is likely to be dominated by the two larger sheets \cite{Singh}.

\begin{figure}
\includegraphics[clip,width=8.7cm]{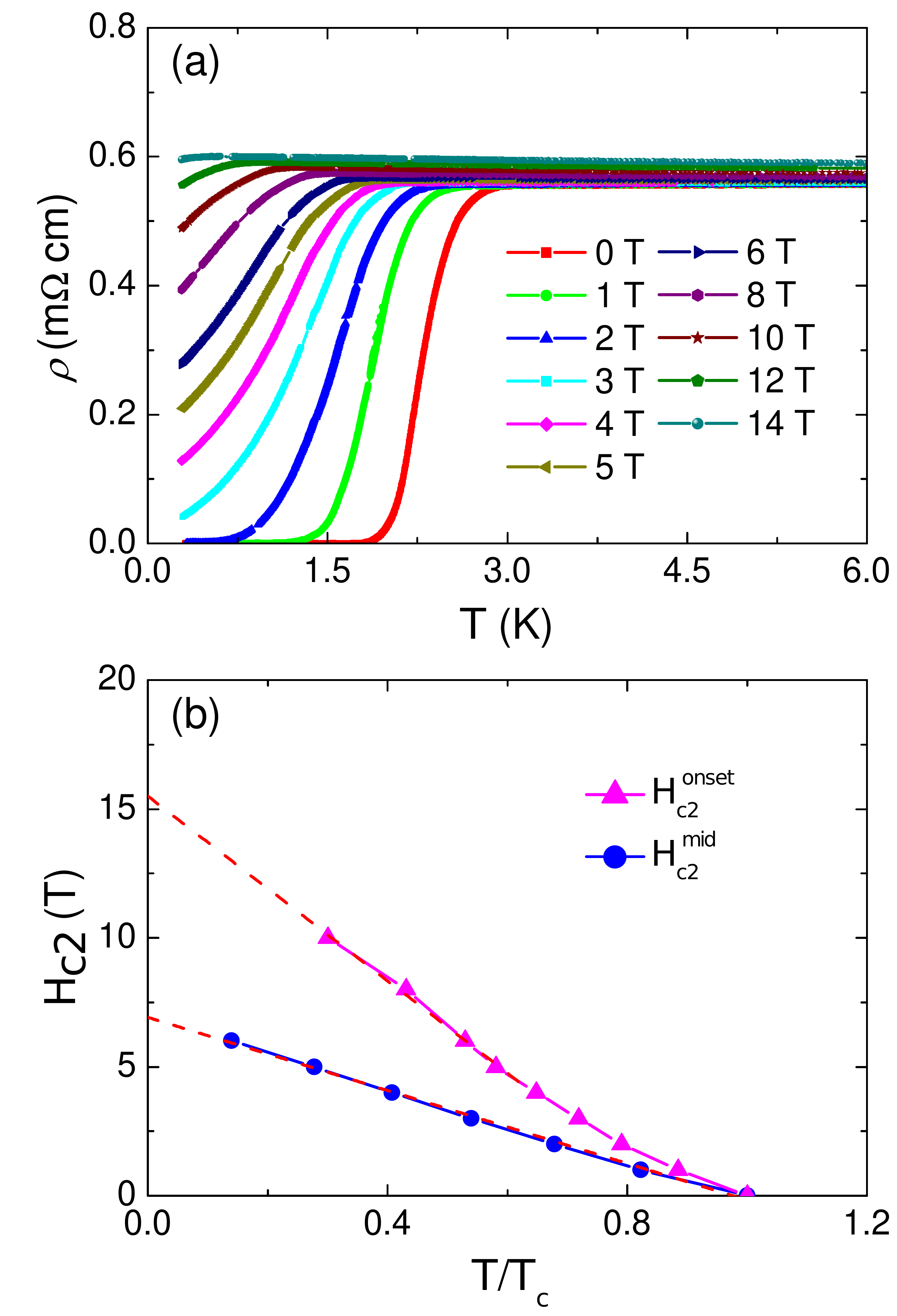}
\caption{(Color online) (a) Low-temperature resistivity $\rho(T)$ of Ta$_2$PdSe$_5$ in various magnetic fields up to 14 T. (b) Reduced temperature $T/T_c$ dependence of the upper critical field $H_{c2}(T)$.  The data points of $H_{c2}^{onset}$ and $H_{c2}^{mid}$ are extracted from each $\rho(T)$ curve in (a) according to the definitions of $T_c^{onset}$ and $T_c^{mid}$.}
\end{figure}

To estimate the coupling strength of Ta$_2$PdSe$_5$, the normalized specific heat jump $\Delta$$C_e/\gamma$$T_c$ is estimated to be about 0.83.
This value is smaller than that expected for weak-coupling BCS superconductors ($\Delta$$C_e$/$\gamma$$T_c$ = 1.43),
which could be attributed to weak coupling strength or a sizeable nonsuperconducting fraction in Ta$_2$PdSe$_5$.
Note that similar value $\Delta$$C_e$/$\gamma$$T_c$ = 0.9 was also reported on the sister compound Nb$_2$PdS$_5$ \cite{Niu,Goyal}.

Figure 4(a) shows the low-temperature resistivity of Ta$_2$PdSe$_5$ in various magnetic fields up to 14 T.
As the magnetic field increases, $T_c$ decreases and the superconducting transition broadens.
At $H$ = 3 T, the zero-resistivity superconductivity has already been suppressed below 0.3 K.
The upper critical field $H_{c2}(T)$ versus reduced temperature $T/T_c$ is shown in Fig. 4(b).
The data points symbolled as $H_{c2}^{onset}$ and $H_{c2}^{mid}$ are extracted from each $\rho(T)$ curve in Fig. 4(a) according to the definitions of $T_c^{onset}$ and $T_c^{mid}$. Both $H_{c2}^{onset}$ and $H_{c2}^{mid}$ increase with decreasing temperature, showing no sign of saturation at the
lowest temperature, in contrast to that observed in a conventional superconductor.
More interestingly, the two curves show roughly linear temperature dependence at low temperature.
We estimate $H_{c2}$(0) $\simeq$ 15.5 T by linearly extrapolating the $H_{c2}^{onset}(T)$ curve to zero temperature. The corresponding ratio of $H_{c2}$(0)/$T_c$ comes out to be 6 T/K. In Table I, we list the $T_c$, $\gamma$, $H_{c2}$(0), and $H_{c2}$(0)/$T_c$ for all four members of the $T_2$Pd$Ch_5$ family \cite{Zhang,Khim,Lu}. Both the linear temperature dependence of $H_{c2}(T)$ and the abnormally high $H_{c2}$(0)/$T_c$ are common features of $T_2$Pd$Ch_5$ \cite{Zhang,Khim,Lu}.

\begin{table}
\centering \caption{Comparison among four $T_2$Pd$Ch_5$ ($T$ = Nb or Ta, $Ch$ = S or Se) compounds, including transition temperature $T_c$, electronic specific heat coefficient $\gamma$, upper critical field $H_{c2}$(0) and the $H_{c2}(0)/T_c$ ratio.  These values are taken from Refs. \cite{Zhang,Khim,Lu}, and this work.}\label{1}
 \begin{center}
  \begin{tabularx}{0.49\textwidth}{p{1.72cm}p{0.55cm}p{2.0cm}p{1.1cm}p{1.2cm}}\hline\hline
  ~~~&$T_c$ & ~~~~~~~$\gamma$ & $H_{c2}$(0)  & $H_{c2}(0)/T_c$   \\
  &(K)& (mJ/mol K$^{2}$) & ~~(T) & ~~(T/K)  \\ \hline
  Nb$_2$Pd$_x$S$_5$ & 6.6 & ~~~~~~~15 & ~~~37   &~~~  5.6   \\
  Nb$_2$Pd$_x$Se$_5$ & 5.0 & ~~~~~~12.8 & ~~~35  &~~~ 5.9    \\
  Ta$_2$Pd$_x$S$_5$ & 5.4 & ~~~~~~27.6  & ~~~31  &~~~ 5.7     \\
  Ta$_2$PdSe$_5$ & 2.5 & ~~~~~~10.3 & ~~15.5  &~~~ 6.2     \\ \hline\hline
  \end{tabularx}
 \end{center}
\end{table}

In Fig. 4(b), a linear fit of the $H_{c2}^{onset}(T)$ data near $T_c$ provides the initial slope $|$d$H_{c2}$/d$T_c$$|$$_{T_c}$ = 4.17 T/K. For one-band BCS superconductors, orbital limiting field $H_{c2}^{orb}$(0) is commonly derived from this slope, and Werthamer-Helfand-Hohenberg (WHH) theory predicts $H_{c2}^{orb}$(0) = -0.69$|$d$H_{c2}$/d$T_c$$|$$_{T_c}$$T_c$ in the dirty limit and $H_{c2}^{orb}$(0) = -0.73$|$d$H_{c2}$/d$T_c$$|$$_{T_c}$$T_c$ in the clean limit. For Ta$_2$PdSe$_5$, the Ginzburg-Landau coherence length $\xi_0$ = 65.8 {\AA} is calculated by using $\xi_0$ = [$\Phi_0$/2$\pi$$H_{c2}$(0)]$^{1/2}$. The electron mean free path $l \simeq$ 24 {\AA} is roughly estimated from $\rho_0$ and $\gamma$ \cite{Khim}. Since $l$ $<$ $\xi_0$, Ta$_2$PdSe$_5$ is close to dirty limit. Therefore, the orbital limit based on one-band WHH formula gives $H_{c2}^{orb}$(0) = -0.69$|$d$H_{c2}$/d$T_c$$|$$_{T_c}$$T_c$ = 7.28 T. This value is higher than the expected weak coupling Pauli limiting field $H_{c2}^{P}$(0) = 1.84$T_c$ = 4.6 T in the case of spin-singlet pairing, but still far below the measured $H_{c2}$(0) value. Such an abnormal $H_{c2}$(0) can not be understood in terms of one-band Ginsburg-Landau theory.

The linear temperature dependence of $H_{c2}(T)$ was previously observed in two-band superconductor MgB$_2$ \cite{Buzea,AGurevich,Ferrando}, and it was explained by Gurevich based on the dirty two-gap Usadel equations \cite{Gurevich,Gurevich1}. Unlike one-gap theory in which $H_{c2}(T)$ has a downward curvature, $H_{c2}(T)$ of a dirty two-gap superconductor can show linear dependence or even upward curvature, depending on the intraband diffusion ratio caused by strong impurity scattering \cite{Gurevich,Gurevich1}. Moreover, the interband scattering can make the $H_{c2}$(0) significantly higher than that estimated from one-gap theory \cite{Mansor}. Therefore, we ascribe the linear temperature dependence of $H_{c2}(T)$ and the abnormally high $H_{c2}$(0)/$T_c$ observed in $T_2$Pd$Ch_5$ family to the multi-band effect. However, one may not ignore the effect of spin-orbital coupling, since large spin-orbit scattering due to Pd deficiency in these compounds can suppress the paramagnetic pair breaking, thus enhance the limit of the upper critical field \cite{Lu}. From the perspective of application, Ta$_2$PdSe$_5$ and the related chalcogenides provide a new family of materials exhibiting high $H_{c2}$, which makes them candidate materials for high-field applications.

\section{Summary}
In summary, a new transition-metal chalcogenide compound Ta$_2$PdSe$_5$ was first synthesized.
Measurements of resistivity, magnetization and specific heat revealed that Ta$_2$PdSe$_5$ is a superconducting material with $T_c$ $\simeq$ 2.5 K.
This compound displays a remarkably high upper critical field $H_{c2}$(0) relative to its superconducting transition temperature $T_c$.
Both the fit of electronic specific heat and the linear temperature dependence of $H_{c2}$ indicate multi-band superconductivity in Ta$_2$PdSe$_5$.

{\it Note added:} Stimulated by the preprint of this work (arXiv:1412.6983), the electronic structure and related properties of Ta$_2$PdSe$_5$ were theoretically obtained from density functional calculations \cite{David}. The Fermi surface has two disconnected sheets \cite{David}, which are consistent with our experimental results.

\acknowledgments
This work is supported by the Natural Science Foundation of China,
the Ministry of Science and Technology of China (National Basic
Research Program No: 2012CB821402 and 2015CB921401),
China Postdoctoral Science Foundation No: 2014M560288, Program for
Professor of Special Appointment (Eastern Scholar) at Shanghai
Institutions of Higher Learning, and STCSM of China (No. 15XD1500200). \\

\end{document}